\documentclass{article}

\usepackage{arxiv}

\usepackage[utf8]{inputenc} 
\usepackage[T1]{fontenc}    
\usepackage{hyperref}       
\usepackage{url}            
\usepackage{booktabs}       
\usepackage{amsfonts}       
\usepackage{nicefrac}       
\usepackage{microtype}      
\usepackage{lipsum}
\usepackage{wrapfig}
\usepackage{lscape}
\usepackage{rotating}
\usepackage{graphicx}
\usepackage{caption}
\usepackage{amsmath}
\usepackage{amstext}
\usepackage{amssymb}
\usepackage{xspace}

\title{METHOD TO CLASSIFY SKIN LESIONS USING DERMOSCOPIC IMAGES}

\author{
	Dusa Sai Charan \\
	Department of Computer Science\\
	Indian Institute of Information Technology,\\ Design and Manufacturing, Kancheepuram\\
	Chennai, Tamil Nadu, India \\
	\texttt{ced16i012@iiitdm.ac.in} \\
	\texttt{hemanthnadipineni@yahoo.in} \\
	\And
	Hemanth Nadipineni\\
	Department of Computer Science\\
	Indian Institute of Information Technology,\\ Design and Manufacturing, Kancheepuram\\
	Chennai, Tamil Nadu, India \\
	\texttt{ced16i024@iiitdm.ac.in} \\
	\And
	Subin Sahayam \\
	Department of Computer Science\\
	Indian Institute of Information Technology,\\ Design and Manufacturing, Kancheepuram\\
	Chennai, Tamil Nadu, India \\
	\texttt{coe18d001@iiitdm.ac.in} \\
	\And
	Umarani Jayaraman \\
	Department of Computer Science\\
	Indian Institute of Information Technology,\\ Design and Manufacturing, Kancheepuram\\
	Chennai, Tamil Nadu, India \\
	\texttt{umarani@iiitdm.ac.in} \\
}
\begin{document}
\maketitle
\vspace{10mm}
\begin{abstract}
Skin cancer is the most common cancer in the existing world constituting one-third of the cancer cases. Benign skin cancers are not fatal, can be cured with proper medication. But it is not the same as the malignant skin cancers. In the case of malignant melanoma, in its peak stage, the maximum life expectancy is less than or equal to 5 years. But, it can be cured if detected in early stages. Though there are numerous clinical procedures, the accuracy of diagnosis falls between 49\% to 81\% and is time-consuming. So, dermoscopy has been brought into the picture. It helped in increasing the accuracy of diagnosis but could not demolish the error-prone behaviour. A quick and less error-prone solution is needed to diagnose this majorly growing skin cancer. This project deals with the usage of deep learning in skin lesion classification. In this project, an automated model for skin lesion classification using dermoscopic images has been developed with CNN(Convolution Neural Networks) as a training model. Convolution neural networks are known for capturing features of an image. So, they are preferred in analyzing medical images to find the characteristics that drive the model towards success. Techniques like data augmentation for tackling class imbalance, segmentation for focusing on the region of interest and 10-fold cross-validation to make the model robust have been brought into the picture. This project also includes usage of certain preprocessing techniques like brightening the images using piece-wise linear transformation function, grayscale conversion of the image, resize the image. This project throws a set of valuable insights on how the accuracy of the model hikes with the bringing of new input strategies, preprocessing techniques. The best accuracy this model could achieve is 0.886. The implementation of the proposed work is available in \footnote[1]{Github Source Code: \href{https://github.com/callmehetch/SkinLesion/blob/master/3b\_unet.ipynb}{Melanoma Classification for Two Path Convolutional Neural Network}}
 
\end{abstract}

\section{Introduction}
\hspace{5em}
Skin lesion refers to the abnormality on the skin. Skin lesions may be of type cancerous, allergic, etc. Among these, cancer-causing skin lesions are hazardous to health. Some forms of these cancerous lesions are deadly. Melanoma is considered to be the one with a high mortality rate of 8\% among the cancerous lesions. The occurrence rate of melanoma is increasing day by day \cite{kopf1982rising}.

Skin cancers are classified majorly into two types. They are \begin{itemize}
    \item Melanoma Skin Cancers (MSC),
    \item Non-Melanoma Skin Cancers (NMSC).
\end{itemize}

In total, there are 8 types of cancerous skin lesions and are classified as follows :
\begin{figure}[!htb]
\centering
  \includegraphics[scale = 0.34]{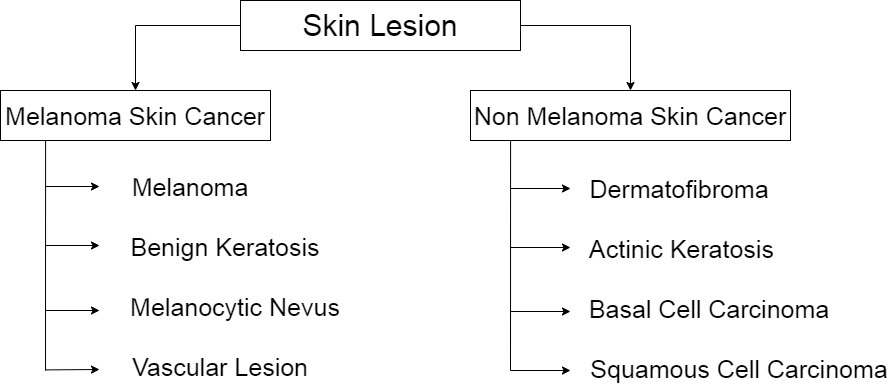}
  \caption[width = 0.1cm]{Classification of skin lesions}
\end{figure}

Clinically, skin cancers are diagnosed based on ABCDE rule \cite{friedman1985early}. Where,
\begin{itemize}
    \item A - Asymmetry
    \item B - Border irregularity
    \item C - Color of lesion
    \item D - Diameter of lesion
    \item E - Enlarging lesion.
\end{itemize}

This above diagnosis can be done through the naked eye. But most skin cancers mimic each other with ABCDE properties. So, there is a possibility of making an error. And there are other clinical procedures like biopsy, etc and are error-prone.  On average, the accuracy of the specialists ranges between 49\% to 81\%, with one-third of melanomas are wrongly predicted as benign lesions \cite{mackenzie1998melanoma, grin1990accuracy, miller1992accurate, morton1998clinical, lindelof1994accuracy}. So dermoscopy came into existence. 

Dermoscopy is one of the most important techniques to examine skin lesions and can capture high-resolution images of the skin escaping the interruption of surface reflections. Dermoscopy helped in making more accurate decisions compared to traditional clinical procedures \cite{argenziano2006dermoscopy}. 

Dermoscopy increased the sensitivity, specificity but could not eradicate the probability of making errors and all the above clinical procedures are time-consuming, requires human efforts. 
Early detection of melanoma can save the lives of people. So, a technique is to be created which is fast and less error-prone.
  
On the other hand, machine learning, especially deep learning-based algorithms have become a methodology of choice for analyzing medical images.  A deep convolution algorithm can be more objective, accurate, and reproducible when it has been well trained. So, this related study lead to the conclusion that developing an automated model for skin cancer classification using dermoscopic images can aid us in taking quicker, error-less decisions. 

This project deals with developing an automated model for skin lesion classification involves the creation of different models and usage of different strategies. The dataset is class imbalanced, so data augmentation is done using affine transformations( rotate, scale, etc) to generate sufficient numbers of images to tackle class imbalance. Skin lesion images have been segmented to the lesion part so that the model is made to concentrate more on the lesion region and exclude the surrounding healthy skin. 10-fold cross-validation has been brought into the picture. Cross-validation makes the model robust. The accuracy will be the mean of the accuracy of all the folds of data. 

\subsection{Motivation}
\begin{itemize}
    \item Manual interpretation demands high time consumption and is highly prone to mistakes.
\item Early detection and treatment can save more than 95\% of people.
\item Automated skin lesion classification model makes the work easier and faster.

\end{itemize}

\subsection{Objectives of the work}
\begin{itemize}
    \item To propose a deep learning model to automatically classify skin lesions.
   \item To validate the proposed model for robustness.
   
\end{itemize}

\section{Literature Work} 

\label{ChapterX} 

Initially, the major work was done on developing an automated model for diagnosing lesions based on ABCDE properties. In these models \cite{kasmi2016classification, abbas2013melanoma, ramteke2013abcd}, they converted these ABCDE properties of each image into numerical figures, generated a cumulative score by adding all the numerical figures of each property and diagnose the images based on the cumulative score. There is another developed on a seven-point checked checklist \cite{di2010automatic}, considered to be an improvisation of ABCDE properties. 


Later, the work was done on the classification of skin lesion images using Raman spectroscopy. Raman spectroscopy is a noninvasive optical procedure fit for estimating vibrational methods of biomolecules inside practical tissues. In this study, they \cite{lieber2008vivo, lui2012real} tested the application of an integrated Raman spectroscopy real-time framework for the in vivo diagnosis of skin cancer
 
In recent times, the major work was carried out on developing models based on CNN variants (ResNet, EfficientNets, SENet, so on) for classifying skin lesions using dermoscopic images \cite{gessert2020skin, 04cook1992ductile, argenziano2000interactive}. They preprocessed the images to eliminate the not interesting regions like eliminating dark circles in the images around the lesion region, contrast balancing, etc. They used data augmentation, random cropping strategies for tackling class imbalance. 5-fold cross-validation, Ensembling strategy for making the model robust.

Let's have a broader look at these three models as they include state-of-the-art CNN architectures, intriguing input strategies, and modern robust training techniques.

The present models are evaluated on ISIC 2019 dataset
\subsection{Input and Expected Output}
The input dataset is ISIC 2019 dataset and it consists of dermoscopic images related to '8'
 different skin lesions. The expected output of the images is a one-hot vector of size '8'
 
\subsection{Preprocessing}
There are certain images in which lesion is surrounded by black color. So this part of the image with pixel value '0' place no role in learning the characteristics of the images. So, the darker parts of the image area to be removed. This helps in two ways - i) Helps us concentrate more on a region of interest. ii) Decreasing the size of the image lessens the computation time.

\begin{figure}[!htb]
\minipage{1\textwidth}
  \includegraphics[scale = 1]{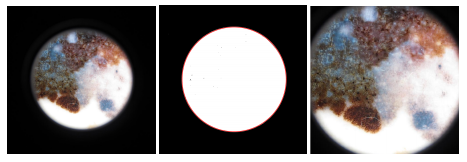}
\endminipage\hfill
\end{figure}

This task is achieved in the following steps: i)Take the image which is surrounded completely by black color. ii) Find a threshold which highlights(make pixel value as '1') the lesion region and darkens(make pixel value as '0') the surrounded black region when the value greater than equal to the threshold is put as '1' and rest as '0'. iii) Find the major axis endpoints and minor axis endpoints of the highlighted regions. extract the rectangular region whose length is equal to the length of the major axis, breadth is equal to the length of the minor axis.

The color contrast of all the images differ a lot so apply Minkowski norm with p = 6 to make the contrast of all the images to as similar as possible 

$$d(x,y) = (\sum_{i=0}^{n-1}|x_{i} - y_{i}|^{p})^{1/p}$$

\section{Deep Learning Models}
\subsection{Architecture}
These papers involved state-of-the-art deep CNN architectures for training. They are i) EfficientNets: This family of models includes eight different models that are structurally similar and are pretrained on the ImageNet dataset. ii)SENet154, iii) ResNet.

The input sizes of all these networks are different. The depth of a network increases with the increase in the size of the input image. 

The count of images of each class is not the same. When a model is trained with this unequal count dataset, the model gets biased in classifying the images. The model may end up predicting the given image as the class with high image count as it learned more characteristics related to that class. So this data imbalance should be tackled to generalize your model. To augment the data.

\subsection{Data Augmentation}
The data augmentation is done through affine transformations and it involves: i) Random brightness, ii) contrast changes, iii) random flipping, iv) random rotation, v) random scaling, vi) random shear, vii) CutOut - This involves creating holes of different sizes on the images i.e. technically making a random portion of image inactive.

The additional data of '995' healthy dermoscopic images from the 7-point dataset are also added to the dataset. The data augmentation technique helps in creating additional so that training dataset can be balanced and training the model with wide varieties of skin lesion image variations. 

\subsection{Input Strategy}
The complete training dataset is segregated into five folds. Each fold consists of images related to one particular class. Two ways of passing inputs have been followed: i)Same-sized cropping: Patches of equal sizes are taken from the preprocessed images and are trained. ii) Random-sized cropping: Patches of random sizes are taken from the preprocesses images and are scaled to the required size are trained.

\subsubsection{Training}
All these models are trained using 'Adam' optimizers, all the other parameters like batch\_size and learning are adjusted based on the GPU memory requirements of each architecture.

Internal evaluation is based on the mean values of sensitivities.
$$S = 1/C \sum_{i=1}^{C} TP_{i} / Tp_{i} + FN_{i}$$

where,\\
TP = True positives, 
FN = Fales negatives, 
C = number of classes.

\subsection{Ensembling}
They create large ensemble out of all the trained models. An optimal subset of models is selected based on cross-validation performance. Top configurations of our choice can be selected and used for prediction.


\subsection{Literature Table}
\begin{table}[h!]
\centering
\resizebox{\textwidth}{!}{\begin{tabular}{|l|l|l|l|}
\hline
Title                                                                                                                                              & Authors                                                                                                                          & Dataset                                                                                           & Target Classes                                                                                                                                                                                   \\ \hline
\begin{tabular}[c]{@{}l@{}}Classification of malignant\\ melanoma and benign skin\\ lesions:  implementation of\\ automatic abcd rule\end{tabular} & \begin{tabular}[c]{@{}l@{}}R. Kasmi,\\ K. Mokrani\end{tabular}                                                                   & EDRA \cite{kasmi2016classification}                                                                                      & \begin{tabular}[c]{@{}l@{}}Melanoma, \\ benign skin \\ lesions\end{tabular}                                                                                                                      \\ \hline
\begin{tabular}[c]{@{}l@{}}Melanoma recognition\\ framework based on\\ expert definition of\\ abcd for dermoscopic images\end{tabular}             & \begin{tabular}[c]{@{}l@{}}Q. Abbas,\\ M. Emre Celebi,\\ I. F. Garcia,\\ W. Ahmad.\end{tabular}                                  & EDRA \cite{abbas2013melanoma}                                                                                     & Melanoma                                                                                                                                                                                         \\ \hline
\begin{tabular}[c]{@{}l@{}}Multi-Category Skin Lesion \\ Diagnosis Using Dermoscopy\\ Images and Deep CNN\end{tabular}                             & \begin{tabular}[c]{@{}l@{}}Steven Zhou1, \\ Yixin Zhuang, \\ Rusong Meng\end{tabular}                                            & ISIC 2019                                                                                & \begin{tabular}[c]{@{}l@{}}Melanoma, Melanocytic Nevus\\ Basal Cell Carcinoma, Actinic Keratosis, \\ \\ Benign Keratosis, Dibrofitoma,\\ Vascular Lesion,\\ Squamous cell Carcinoma\end{tabular} \\ \hline
\begin{tabular}[c]{@{}l@{}}Automatic diagnosis of\\ melanoma:  a software\\ system based on the\\ 7-point check-list\end{tabular}                  & \begin{tabular}[c]{@{}l@{}}G. Di Leo,\\ A. Paolillo,\\ P. Sommella,\\ G. Fabbrocini\end{tabular}                                 & public \cite{di2010automatic}                                                                                   & Melanoma                                                                                                                                                                                         \\ \hline
\begin{tabular}[c]{@{}l@{}}In vivo nonmelanoma skin\\ cancer diagnosis using\\ raman microspectroscopy\end{tabular}                                & \begin{tabular}[c]{@{}l@{}}C. A. Lieber,\\ S. K. Majumder,\\ D. L. Ellis,\\ D. D. Billheimer,\\ A. Mahadevan-Jansen\end{tabular} & IRB(patients) \cite{lieber2008vivo}                                                                                   & \begin{tabular}[c]{@{}l@{}}Melanoma, Melanocytic Nevus\\ Basal Cell Carcinoma, Actinic Keratosis, \\ Benign Keratosis, Dibrofitoma,\\ Vascular Lesion,\\ Squamous cell Carcinoma\end{tabular}    \\ \hline
\begin{tabular}[c]{@{}l@{}}Real-time raman\\ spectroscopy for in vivo\\ skin cancer diagnosis\end{tabular}                                         & \begin{tabular}[c]{@{}l@{}}H. Lui, J. Zhao,\\ D. McLean,,\\ H. Zeng\end{tabular}                                                 & \begin{tabular}[c]{@{}l@{}}Vancouver General\\ Hospital Skin Care\\ Centre(patients) \cite{lui2012real} \end{tabular} & \begin{tabular}[c]{@{}l@{}}Melanoma, Melanocytic Nevus\\ Basal Cell Carcinoma, Actinic Keratosis, \\ Benign Keratosis, Dibrofitoma,\\ Vascular Lesion,\\ Squamous cell Carcinoma\end{tabular}    \\ \hline
\begin{tabular}[c]{@{}l@{}}Skin lesion classification\\ using ensembles of multi-\\ resolution efficientnets\\ with meta data\end{tabular}         & \begin{tabular}[c]{@{}l@{}}N. Gessert,\\ M. Nielsen,\\ M. Shaikh, R. Werner,\\ A. Schlaefe\end{tabular}                          & ISIC 2019 \cite{gessert2020skin}                                                                                & \begin{tabular}[c]{@{}l@{}}Melanoma, Melanocytic Nevus\\ Basal Cell Carcinoma, Actinic Keratosis, \\ Benign Keratosis, Dibrofitoma,\\ Vascular Lesion,\\ Squamous cell Carcinoma\end{tabular}    \\ \hline
\end{tabular}}
\end{table}
\newpage
\section{Proposed method} 

\label{ChapterX} 

This project involves development of three models -
\begin{itemize}
    \item Model - 1 
    \item Model - 2 :
    \begin{itemize}
        \item CNN Model - 1 - One Path
        \item CNN Model - 2 - Dual Path
    \end{itemize}
\end{itemize}
where, model - 2 involves two CNN Models. The 'Model-1' takes the original images as input, 'Model-2' with 'CNN Model - 1' takes image formed with the multiplication of original image and segmented mask as an input ans 'Model - 2' with 'CNN Model - 2' takes both original image and segmented mask as inputs.
There is the usage of different input strategies, preprocessing techniques, and training techniques. The detailed description is as follows : 
\subsection{Input and Expected Output}
The input is ISIC 2019 dataset and the expected output is 'one-hot vector' of size '8'as explained above.
\begin{figure}[!htb]
\minipage{1\textwidth}
  \includegraphics[scale = 0.55]{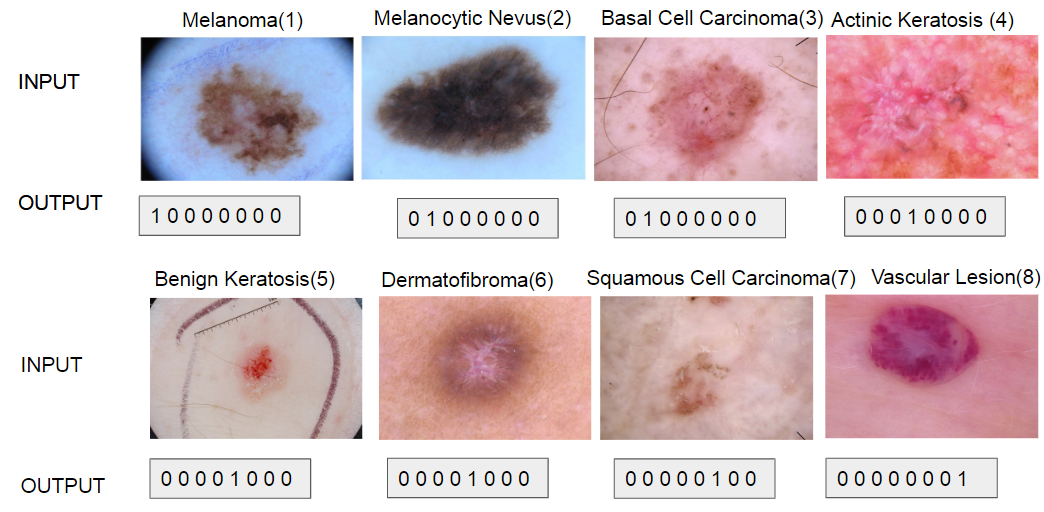}
  \caption[width = 0.1cm]{Input and expected ouput of the model developed}
\endminipage\hfill
\end{figure}

The dataset is an accumulation of 3 different datasets. 
\begin{itemize}
    \item Ham\_1000 (600 x 450) pixels.
    \item BCN\_2000 (1024 x 1024) pixels.
    \item MSK (1024 x 1024) pixels.
\end{itemize}

The total count of images is 25331. The count of images for each class is unequal which leads to class imbalance. Each class images’ resolutions differ from other class images. 

MLN        MCN        BCC        AK        BK       DF        VL        SCC\\
    4522    12875    3323    867    2624    239    253    628\\
    
Challenges faced: i) Class imbalance, ii) varying contrast of images, iii) Improper images(lesion color is similar to the skin color), iv) Different resolutions for images.

\newpage

\subsection{Classification Model - 1}
\vspace{5mm}
The 'Model - 1' involves simple preprocessing of the input image, then pass the preprocessed image as an input to the CNN model, train the model, and start classifying the images.
\vspace{15mm}
\begin{figure}[!htb]
\minipage{1\textwidth}
  \includegraphics[scale = 0.4]{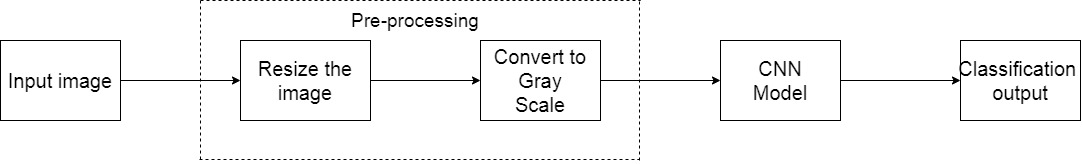}
  \caption[width = 0.1cm]{Architecture of Classifcation 'Model - 1' }
\endminipage\hfill
\end{figure}

This model is used to study the effectiveness of CNN in capturing features of an image for classification. The major concentration is on CNN and it’s variants as they are good at capturing neighbourhood properties in an image which is essential in differentiating one skin cancer image from another image.
\vspace{15mm}

\subsubsection{Preprocessing}
All the input images are of various sizes so to pass these images as input to the model developed they should be of the same size. So, in this phase, all the images are scaled down to the size 512 x 512 pixels. These images are further converted into grayscale. This is from our intuition that skin lesion classification does not depend on the color of the lesion.
\vspace{15mm}

\begin{figure}[!htb]
\minipage{1\textwidth}
  \includegraphics[scale = 0.9]{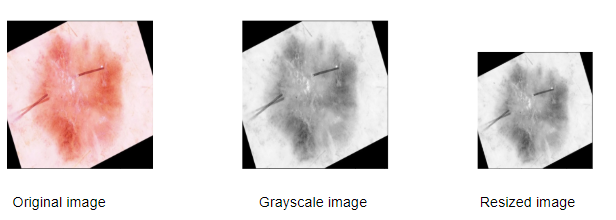}
  \caption[width = 0.1cm]{shows how an input image is preprocessed to the required form}
\endminipage\hfill
\end{figure}
\newpage
\subsubsection{CNN Architecture of Classification 'Model - 1'}
This is the architecture of the developed CNN Model :\\
\begin{figure}[!htb]
\minipage{1\textwidth}
  \includegraphics[scale = 0.4]{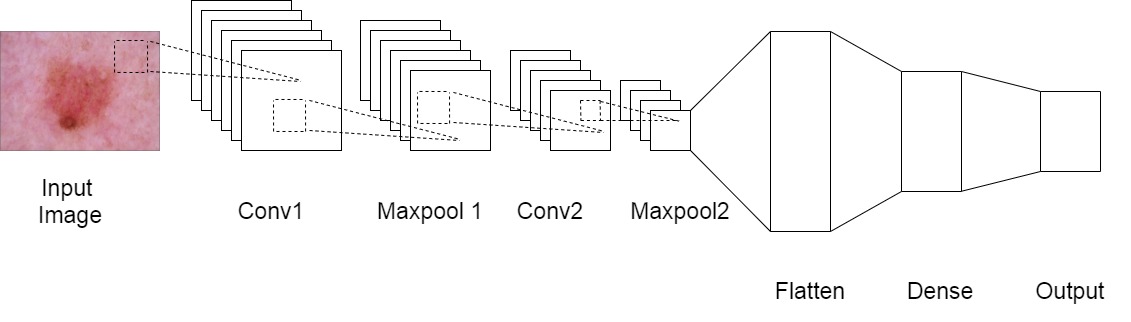}
  \caption[width = 0.1cm]{CNN architecture of Classification 'Model - 1'}
\endminipage\hfill
\end{figure}\\
The size of the input image is '512 x 512' pixels and the kernel size is '3 x 3'. In the convolution layer - 1, '64' filters are used which results in the creation of '64' feature maps and size of the output image shrinks to '510 x 510' pixels. Then this output is passed as an input to the max pool layer - 1 with stride '2' for the filters, which results in the creation of '64' feature maps whereas the size of the output images shrinks to the size '255 x 255'. Similarly, the number of feature maps at convolution layer - 2 is '32', the output image size is '253 x 253'. The number of feature maps at max pool layer - 2 is 32, where the size of the output image is '121 x 121'. At flattening stage, the count of neurons is 121 x 121 x 32 = '468512'. The number of a neuron at Dense - 32 layer is '32'. The number of neurons at the output layer is '8'.
\vspace{15mm}
The model is trained on 90 percent of images with optimizer ‘Adams’ for ‘20’ epochs. The activation function at convolution and dense layers are 'Relu'.
$$x = max(0,x)$$

The activation function at output layer is 'softmax'.
$$z_{i} = e^{x_{i}} / \sum_{j=1}^{K} e^{x_{j}}$$

The loss function used is 'categorical cross\_entropy'
$$L(\hat{y},y) = -1/N\sum_{i}^{N}[y_{i}\log \hat{y}_{i} + (1-y_{i})\log(1- \hat{y}_{i})]$$

\vspace{15mm}
The accuracy is ~0.204. After going through past work done in the field and clinical procedure for the diagnosis, there are some important issues where model-1 least stressed.

The following can be considered as the drawbacks of the above model: i) Not focussing more on regions of interest, ii) item Color plays a major role in differentiation, iii) No tackling for class imbalance.
\newpage
\subsection{Classification Model-2}

This model includes segmentation which helps us in extracting the region of interest and helps us in concentrating more on the region of interest. The input for the model is the same as the input for the model-1. 

\begin{figure}[!htb]
\minipage{1\textwidth}
  \includegraphics[scale = 0.6]{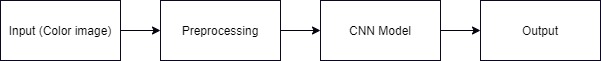}
  \caption[width = 0.1cm]{Architecture of Classification Model - 2}
\endminipage\hfill
\end{figure}
\vspace{15mm}
\subsubsection{Preprocessing}

The required lesion region of interest is segmented out from the given dermoscopic images
\begin{figure}[!htb]
\minipage{1\textwidth}
  \includegraphics[scale = 0.5]{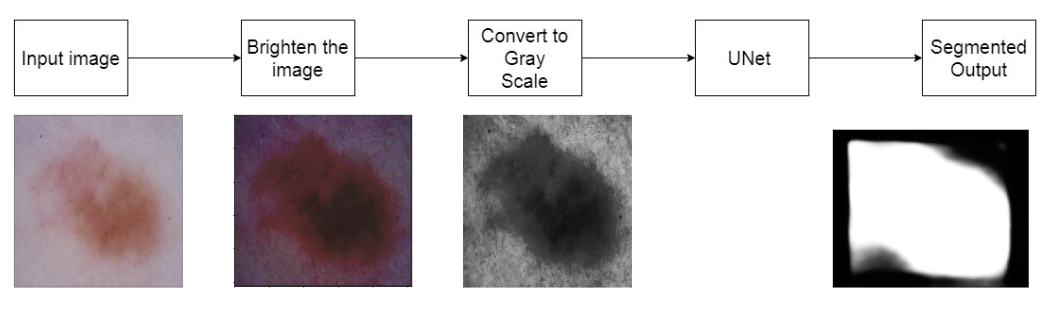}
\caption[width = 0.1cm]{shows how an input image is preprocessed to the required state}
\endminipage\hfill
\end{figure}

Steps involved are :
\vspace{15mm}
\begin{itemize}
    \item \textbf{Brighten the image :} Color of some class images is so light that there is no distinction between the colors of skin and the lesion. This task is achieved using the ‘piece-wise linear transformation’ function.
    
    \begin{figure}[!htb]
    \minipage{1\textwidth}
      \includegraphics[scale = 0.2]{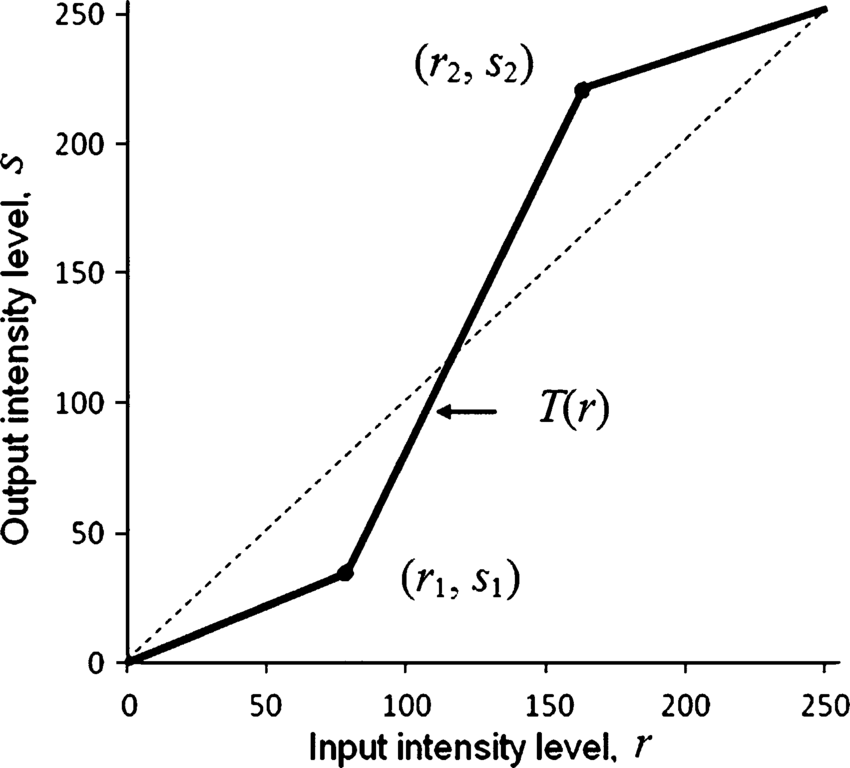}
      \caption[width = 0.1cm]{Piece-wise linear function}
    \endminipage\hfill
    \end{figure}
    
    \newpage
    Where,

    \begin{itemize}
    \item (r1,s1) , (r2,s2) as parameters stretches intensity levels,
    \item Decreases intensity of dark pixels,
    \item Increases intensity of bright pixels.
    \end{itemize}
    \vspace{15mm}
    \textbf{Logic for piece-wise linear transformation :}
    
    def pixelVal(pix, r1, s1, r2, s2):\\ 
        if (0 <= pix and pix <= r1):\\ 
            return (s1 / r1)*pix\\
        elif (r1 < pix and pix <= r2):\\ 
            return ((s2 - s1)/(r2 - r1)) * (pix - r1) + s1\\ 
        else:\\ 
            return ((255 - s2)/(255 - r2)) * (pix - r2) + s2\\ 
    
    \item \textbf{Grayscale the image :} Convert image to grayscale, this makes the lesion region to stand out with high brightness.
  
    \item \textbf{U-net Architecture :} U-net,which is a CNN variant is the state-of-the-art architecture for segmentation.
    \vspace{10mm}
    \begin{figure}[!htb]
    \minipage{1\textwidth}
      \includegraphics[scale = 0.3]{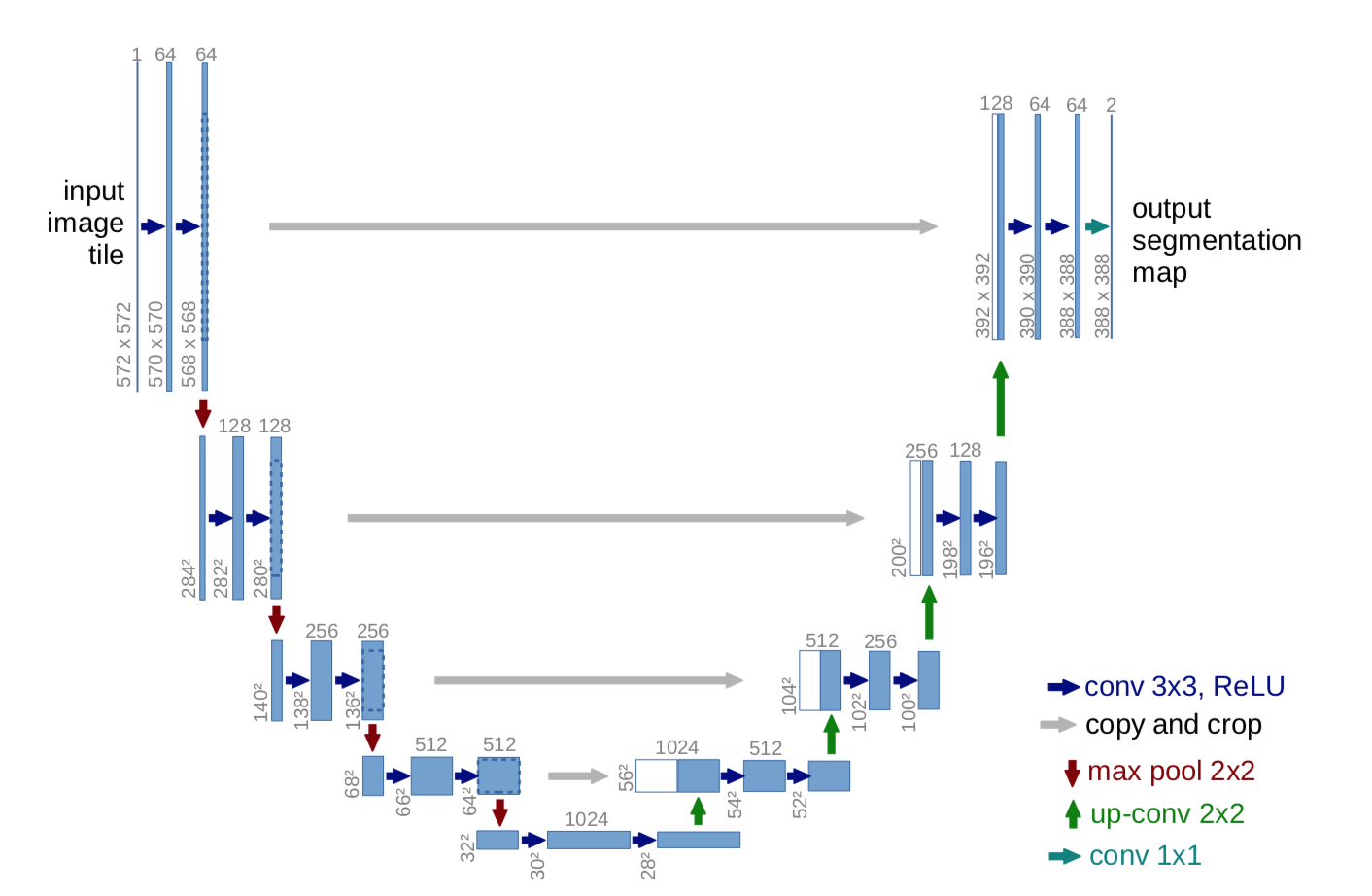}
      \caption[width = 0.1cm]{Architecture of U-Net}
    \endminipage\hfill
    \end{figure}
    
\end{itemize}
\vspace{20mm}
\textbf{Sample Outputs :}
\begin{figure}[!htb]
\minipage{1\textwidth}
  \includegraphics[scale = 0.5]{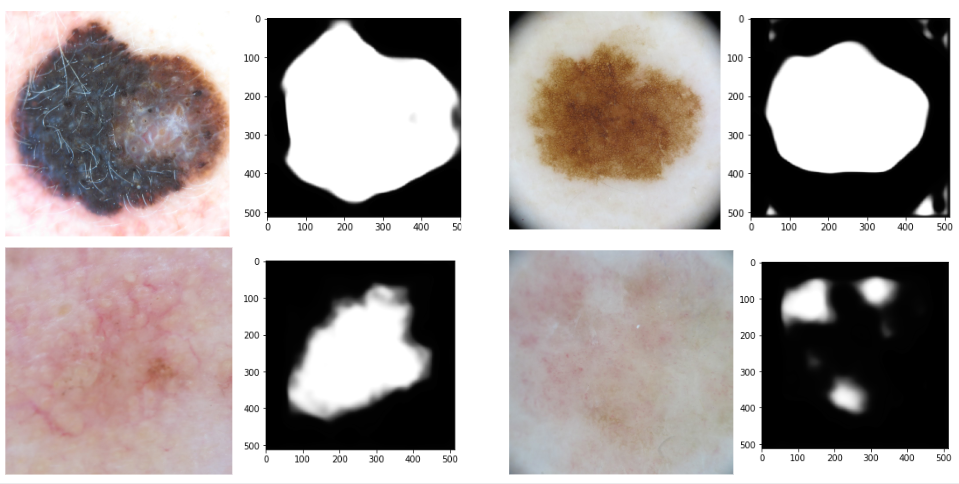}
  \caption[width = 0.1cm]{Input images and their predicted segmentation masks}
\endminipage\hfill
\end{figure}

\subsection{Tackling Data imbalance}
The count of images for each class is -

\begin{table}[h!]
\begin{tabular}{|l|l|l|l|l|l|l|l|}
\hline
MLN  & MCN   & BCC  & AK  & BK   & DF  & VL  & SCC \\ \hline
4522 & 12875 & 3323 & 867 & 2624 & 239 & 253 & 628 \\ \hline
\end{tabular}
\end{table}

The least size is ‘239’. So picking 239 images from each class, training the model with a dataset of size ‘239 * 8’ images will not be robust. So, additional data is to be created, and for that to happen ‘Data augmentation’ should be brought into the picture. 
The list of the classes with count lesser than 2000:
AK, DF, VL, SCC

So additional data has been created for the above-mentioned classes using affine transformations. Affine transformations include rotation, cropping, scaling, flipping(horizontal and vertical), shearing, changing contrast.

\subsection{K-Fold Cross Validation}
K-fold cross-validation technique makes the model robust. This model is made robust through 10-fold cross-validation in which 1-fold is for validation and 9-other folds are for training.   

So, the plan is to train the model on the dataset of 16000 images(2000 images from each class). The model is to be trained for 10 times by giving a chance for each fold to act as a validation data set. Find the best model at each time of training and average the accuracies.
The mean accuracy will be the accuracy of the model.  

\subsection{CNN Model}
The CNN model section explains about two developed CNN models. Of which one is one pathed and another one is dual pathed.
\vspace{5mm}
\subsubsection{CNN Model - 1(One path)}
This CNN model -1 takes a multiplied image as an input. 
The multiplied image is created by multiplying the original image with its corresponding segmentation output. 
\newpage
\begin{figure}[!htb]
\minipage{1\textwidth}
  \includegraphics[scale = 0.5]{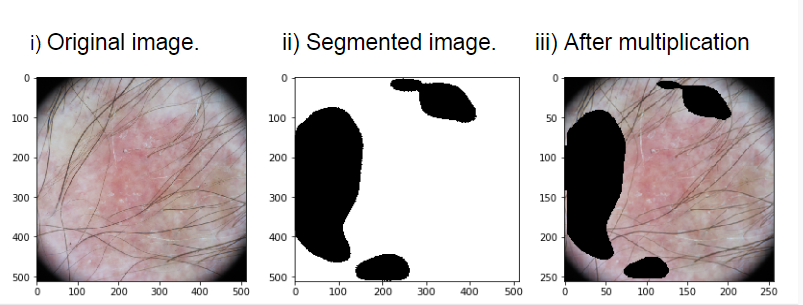}
  \caption[width = 0.1cm]{Resultant mutliplied image}
\endminipage\hfill
\end{figure}

The architecture of CNN Model - 1 that makes use of this multiplied image.
\begin{figure}[!htb]
\minipage{1\textwidth}
  \includegraphics[scale = 0.5]{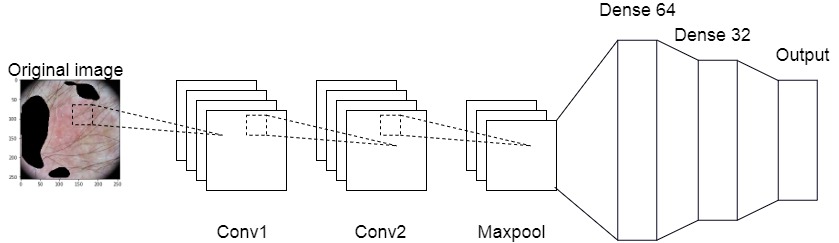}
  \caption[width = 0.1cm]{Architecture of CNN Model - 1(One path)}
\endminipage\hfill
\end{figure}

The size of the input image is '256 x 256 x 3' pixels and the kernel size is '3 x 3'. In the convolution layer - 1, '32' filters are used which results in the creation of '32' feature maps and size of the output image shrinks to '254 x 254' pixels. Then this output is passed as an input to the convolution layer - 2, which results in the creation of '64' feature maps whereas the size of the output images shrinks to the size '252 x 252'. this output is passed as an input to the max pooling layer with stride - '2', which results in the creation of '64' feature maps whereas the size of the output images shrinks to the size '126 x 126'. At flattening stage, the count of neurons is 126 x 126 x 64 = '1,016,064'. The number of neurons at Dense - 64 layer is '64'. The number of neurons at Dense - 32 layer is '32'. The number of neurons at the output layer is '8'.

\vspace{10mm}
\subsubsection{CNN Model - 2(Two path)} 
This model takes both the original image and segmented images as inputs, there are parallel similar architectures for processing these inputs into the next levels. These images are combined at the flattening stage and are connected to the dense levels.
\newpage
\begin{figure}[!htb]
\minipage{1\textwidth}
  \includegraphics[scale = 0.5]{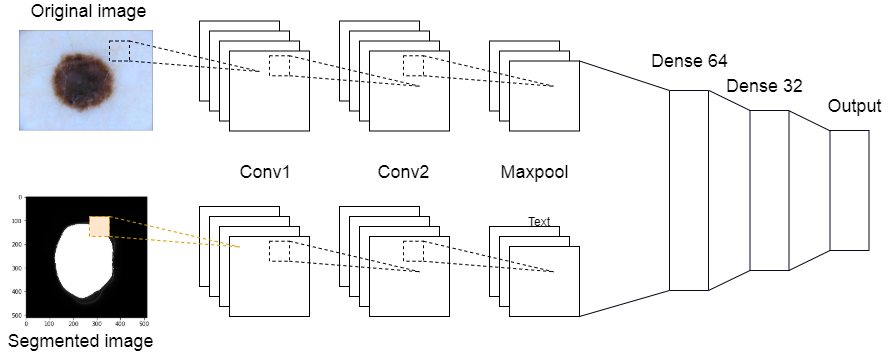}
  \caption[width = 0.1cm]{Architecture of CNN Model - 2(Two Path)}
\endminipage\hfill
\end{figure}

The output is a one-hot vector which points to the predicted class with value ‘1’.

The size of the input image passed to (path - 1) is '256 x 256 x 3' pixels and the kernel size is '3 x 3'. In the convolution layer - 1, '32' filters are used which results in the creation of '32' feature maps and size of the output image shrinks to '254 x 254' pixels. Then this output is passed as an input to the convolution layer - 2, which results in the creation of '64' feature maps whereas the size of the output images shrinks to the size '252 x 252'. this output is passed as an input to the max pooling layer with stride - '2', which results in the creation of '64' feature maps whereas the size of the output images shrinks to the size '126 x 126'. At flattening stage, the count of neurons is 126 x 126 x 64 = '1,016,064'. 

At the flattening stage 'path - 1' which takes the original image as input produces '1,016,064' neurons. So, till the flattening stage architecture is the same as 'path - 1'. So the number of neurons contributed by 'path - 2' to the flattening stage is also '1,016, 064'. Therefore, the total count of neurons at the flattening stage is '2,032,128'.

The number of the neuron at Dense - 64 layer is '64'. The number of the neuron at Dense - 32 layer is '32'. The number of neurons at the output layer is '8'.

The model is trained on 9-folds of images each time whereas the other one-fold will acts as a validation data set. The process has to be repeated until each acts as a validation set exactly once. The model is trained with optimizer ‘Adams’ for ‘2’ epochs. The batch\_size is 75. The initialization of weights is Keras default initialization i.e. Xavier uniform initialization. The activation function at convolution and dense layers are 'Relu'.
$$x = max(0,x)$$

The activation function at output layer is 'softmax'.
$$z_{i} = e^{x_{i}} / \sum_{j=1}^{K} e^{x_{j}}$$

The loss function used is 'categorical cross\_entropy'
$$L(\hat{p},p) = -1/N\sum_{i}^{N}[p_{i}\log \hat{p}_{i} + (1-p_{i})\log(1- \hat{p}_{i})]$$

The accuracy of the model is mean of the accuracies of all the k-models.

$$Acc = (\sum_{i=1}^{K}Acc_{i} ) / K$$

\newpage

\section{Results} 

\label{ChapterX} 

The following table describes the accuracy of the models with different ways of attempt.

\begin{table}[h!]
\begin{tabular}{|l|l|l|l|}
\hline
Ways of attempt                                                                                        & Model - 1 & \begin{tabular}[c]{@{}l@{}}Model - 2 with\\ one path -CNN model\end{tabular} & \begin{tabular}[c]{@{}l@{}}Model-2 with\\ two path CNN model\end{tabular} \\ \hline
\begin{tabular}[c]{@{}l@{}}Without\\ Segmentation\end{tabular}                                         & 0.204     & N.A                                                                          & N.A                                                                       \\ \hline
\begin{tabular}[c]{@{}l@{}}With \\ Segmentation\end{tabular}                                           & 0.412     & 0.561                                                                        & 0.576                                                                     \\ \hline
\begin{tabular}[c]{@{}l@{}}With Segmentation,\\ Augmented data\end{tabular}                            & 0.483     & 0.681                                                                        & 0.732                                                                     \\ \hline
\begin{tabular}[c]{@{}l@{}}With Segmentation,\\ Augmented data\\ 10-fold cross validation\end{tabular} & 0.583     & 0.814                                                                        & 0.886                                                                     \\ \hline
\end{tabular}
\end{table}

Accuracy measure  =     TP/( TP + FN)\\
Where,\\
True positive = predicting the true class correctly,\\
False negative = predicting the true class incorrectly.

When K-fold cross-validation is used accuracy measure is not just the accuracy of one model but the mean of the accuracies of the K-models.

So, the accuracy of the model with K-fold cross-validation is 
$$Acc = (\sum_{i=1}^{K}Acc_{i} ) / K$$

This table shows the comparison of this project with the state-of-the-art models in terms of accuracy.

\begin{table}[h!]
\centering
\begin{tabular}{|l|l|}
\hline
Model                                                                                                                                                 & Accuracy \\ \hline
\begin{tabular}[c]{@{}l@{}}Skin lesion classification \\ using loss balancing and ensemble \\ of multi-resolution efficient nets \cite{gessert2020skin} \end{tabular} & 0.926    \\ \hline
\begin{tabular}[c]{@{}l@{}}Multi-Category Skin Lesion \\ Diagnosis using dermoscopic images\\ and deep CNN models \cite{04cook1992ductile} \end{tabular}               & 0.917    \\ \hline
\begin{tabular}[c]{@{}l@{}}Classification Model - 2\\ with two path CNN model \\ (The present project)\end{tabular}                                   & 0.886    \\ \hline
\begin{tabular}[c]{@{}l@{}}ISIC 2019 Skin lesion analysis\\ towards melanoma detection \cite{a}\end{tabular}                                          & 0.851    \\ \hline
\begin{tabular}[c]{@{}l@{}}Skin lesion analysis towards\\ melanoma detection  using siamese\\ network \end{tabular}                           & 0.832    \\ \hline
\end{tabular}
\end{table}

\subsection{Inferences}

The accuracy of 'Model - 1' is less than in all ways compared to 'Model - 2 with one path CNN model' and 'Model - 2 with two path CNN model'. The explanation for this kind of behaviour can be made from the number of neurons involved in the training of the model. The number of neurons at the flattening layer for i) 'Model - 1' is '4,68,512', ii) 'Model - 2 with one path CNN model' is 1,016,064. iii) 'Model - 2 with two path CNN model' is 2,032,128. But there is a significant difference in the number of neurons between 'Model - 2 with one path CNN model' and  'Model - 2 with two path CNN model' whereas there is no significant difference in the accuracy. So, there is threshold between '4,68,512' - '1,016,064' that surges up the accuracy but there is no threshold between '1,016,064' - '2,032,128'. 

There is an improvement in the accuracy of injecting segmentation into the model. This is happening because when the model is trained with an original image without segmentation, the model tends to learn about the unwanted characteristics of lesion classification. From the [fig 3.1] we can observe that size of the skin lesion region is small compared to the image size. So, to concentrate on the region of interest, the unwanted portion is to be eliminated. So when the image the original is multiplied with the segmented mask, the pixel values of an unwanted portion is set to '0'. So they don't involve in training.

There is an improvement in accuracy when data augmentation is involved. This is because the model is to made to learn on multiple skin variations. So this helped model to learn variety characteristics related to skin lesions. Therefore, it conveys that model with data augmentation tend to learn more distinguishing characteristics features rather like a model without data augmentation.

Involving segmentation, data augmentation, and other preprocessing techniques can improve the accuracy of the model. But that model cannot give significant results with the testing data until it is robust. The model trained can become biased towards a certain class. So, to develop a generalized model, it should be trained with a balanced data set. 

\section{Conclusion and Future Work} 

\label{ChapterX} 


\subsection{Conclusion}
In this project of developing an automated model for skin lesion classification, major work has done on improving the accuracy of the model by bringing in new techniques into the model. 

It is clear from the results, that there is an increment in the accuracy of bringing new strategies into the picture. Though the CNN model is two-layered, it is inferring that the effective preprocessing, input, and training strategies can bring a significant change in the model's accuracy. The segmentation model 'U-net' worked well.

\subsection{Future Work}
The developed 'model-2 with CNN mode-1' which is of depth two layers produced significant results. But bringing in the deep CNN models will be very helpful in learning the distinguishing characteristics of the skin lesions from dermoscopic images.

Work extensions - 
\begin{itemize}
    \item Bringing in deep CNN variants (EfficientNets, SENet, ResNet, so on)
    \item Involving ensembling methods can create a significant imapact in improving the accuracy.
\end{itemize}

\bibliographystyle{unsrt}  
\bibliography{references}  

\end{document}